\documentclass{elsart3p}

\usepackage{graphicx}

\usepackage{amssymb}

\begin{document}

\begin{frontmatter}



\title{Vortex core shrinkage in a two gap superconductor: \\ application to MgB$_2$}


\author{S. Graser,}
\author{A. Gumann, \corauthref{cor1}}
\author{T. Dahm, and N. Schopohl}
\corauth[cor1]{Corresponding author.}
\address{Institut f\"ur Theoretische Physik, 
         Universit\"at T\"ubingen, 
         Auf der Morgenstelle 14, D-72076 T\"ubingen, 
         Germany}

\begin{abstract}
As a model for the vortex core in MgB$_2$ we study a two band model
with a clean $\sigma$ band and a dirty $\pi$ band. We present 
calculations of the vortex core size in both bands as a function 
of temperature and show that there exists a Kramer-Pesch effect
in both bands even though only one of the bands is in the
clean limit. We present calculations for different $\pi$ band
diffusivities and coherence lengths.
\end{abstract}

\end{frontmatter}


In a recent work \cite{Gumann} we have studied the vortex core shrinkage
at low temperatures
(Kramer-Pesch effect \cite{KP1}) in a two-gap superconductor.
In a clean single gap superconductor the Kramer-Pesch effect
results in a linear decrease of the vortex core size as
a function of temperature \cite{KP1}. However, in the dirty limit
the size of the vortex core saturates to a finite value of the
order of the coherence length and the Kramer-Pesch effect is
absent. In the two gap superconductor MgB$_2$ we have an
interesting new situation: in high quality MgB$_2$ samples
it is believed that the $\sigma$ bands are in the clean, while the
$\pi$ bands are still in the dirty limit \cite{Quilty,Putti}.
This motivated us to study the Kramer-Pesch effect within a 
mixed model, which was proposed recently by Tanaka et 
al \cite{Tanaka}, consisting of a clean $\sigma$ band and
a dirty $\pi$ band. The surprising result of our study was
that the clean band {\it induces} a Kramer-Pesch effect in the
dirty band \cite{Gumann}, i.e. the vortex core size decreases
linearly as a function of temperature even in the dirty
$\pi$ band. This ``induced'' Kramer-Pesch effect should
be very interesting for experimental verifications of
the Kramer-Pesch effect. Experimentally in MgB$_2$
it is much easier to image the vortex core in the $\pi$ 
band than in the $\sigma$ band by scanning tunneling
microscopy (STM), because tunneling is easier into the
$\pi$ band \cite{Eskildsen}.

In the present work we want to study the influence
of the $\pi$ band coherence length on the
induced Kramer-Pesch effect. The coherence
length in the clean $\sigma$ band is given by
$\xi^{(\sigma)} = \hbar v_{F,\sigma}/\Delta^{(\sigma)}$
and thus does not vary much with the amount
of impurities in the $\pi$ band. However, the
coherence length in the dirty $\pi$ band is given
by $\xi^{(\pi)}=\sqrt{\mathcal{D}^{(\pi)}/ 2\pi T_{c}}$,
where $\mathcal{D}^{(\pi)}$ is the diffusivity in the
$\pi$ band. Thus, the $\pi$ band coherence length
varies with the impurity scattering rate and should
be sample dependent.

For our numerical calculations we use the Riccati method
to solve Eilenberger's equations in the clean band,
while for the dirty band we solve Usadel's equations
in the vicinity of a single vortex.
The gaps in both bands are calculated self-consistently
and the vortex core radii are calculated from the
slope of the gap function at the vortex center as
has been suggested by Hayashi et al \cite{Hayashi}:
\begin{equation}
\left(\xi_{V}^{(\alpha)}\right)^{-1}  =  \left.\frac{\partial\Delta^{(\alpha)}(r)}{\partial r}
\right|_{r=0}\frac{1}{\Delta^{(\alpha)}(r=\infty,\, T)}
\end{equation}
Parameters appropriate for MgB$_2$ have been used
as has been described in our previous work \cite{Gumann}.

In Fig. 1 we show the temperature dependence of the vortex core radius
in both bands relative to the (fixed) $\sigma$ band coherence length
for three different values of the $\pi$ band coherence length.
The vortex core radius in the $\sigma$ band $\xi^{(\sigma)}_V$ is 
only weakly affected
by the change of the $\pi$ band coherence length and shows a clear
linear decrease, similar as in a clean single band superconductor.
However, the vortex core radius in the $\pi$ band $\xi^{(\pi)}_V$
strongly depends
on the $\pi$ band coherence length and its temperature dependence
shows some deviations from the linear behavior. This can
be seen more clearly in the insets, where we are showing the
temperature dependence of the ratio of the vortex core sizes
in the two bands $\xi^{(\pi)}_V/\xi^{(\sigma)}_V$. For a large
$\pi$ band coherence length (lowest panel) this ratio increases
almost linearly with decreasing temperature, while for a short 
$\pi$ band coherence length (upper panel) the $\pi$ band
vortex core size at high temperature mostly follows the
$\sigma$ band vortex core size and then decreases more
slowly at low temperatures. However, it is very clear that in all
three cases the $\pi$ band vortex core size eventually goes
to 0 for $T \rightarrow 0$, even though the $\pi$ band is
in the dirty limit. Note, that the $\pi$ band vortex core
is always larger than the vortex core in the $\sigma$ band,
even in the case where the $\pi$ band coherence length
is much smaller than the $\sigma$ band coherence length.
We attribute this to the fact that
the dirty $\pi$ band cannot support its own vortex core shrinkage
and is mostly induced by the one in the $\sigma$ band.

\begin{figure}[t]
  \begin{center}
    \includegraphics[width=0.98\columnwidth,angle=0]{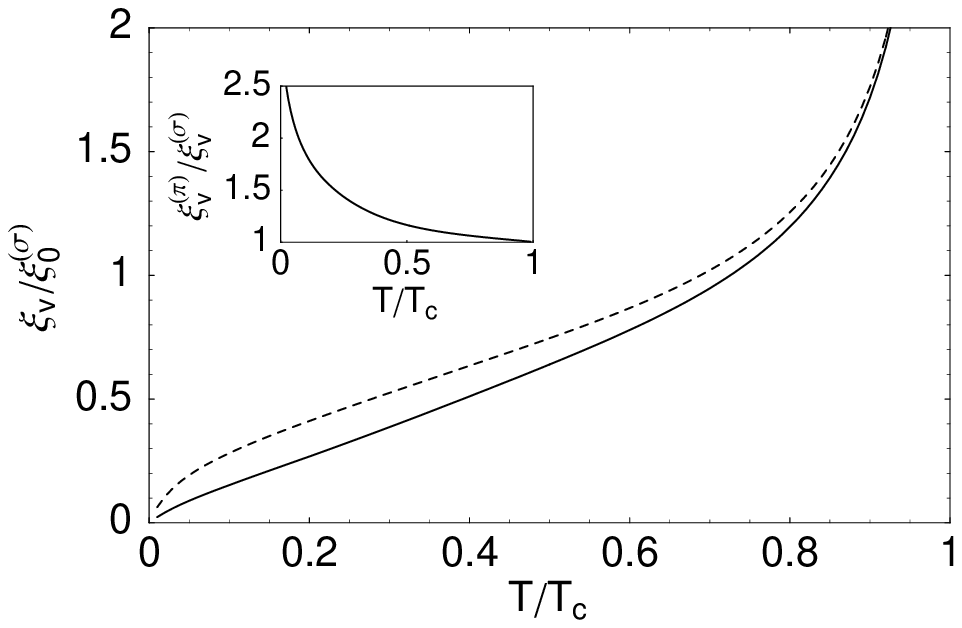}
\vspace*{0.5ex}

    \includegraphics[width=0.95\columnwidth,angle=0]{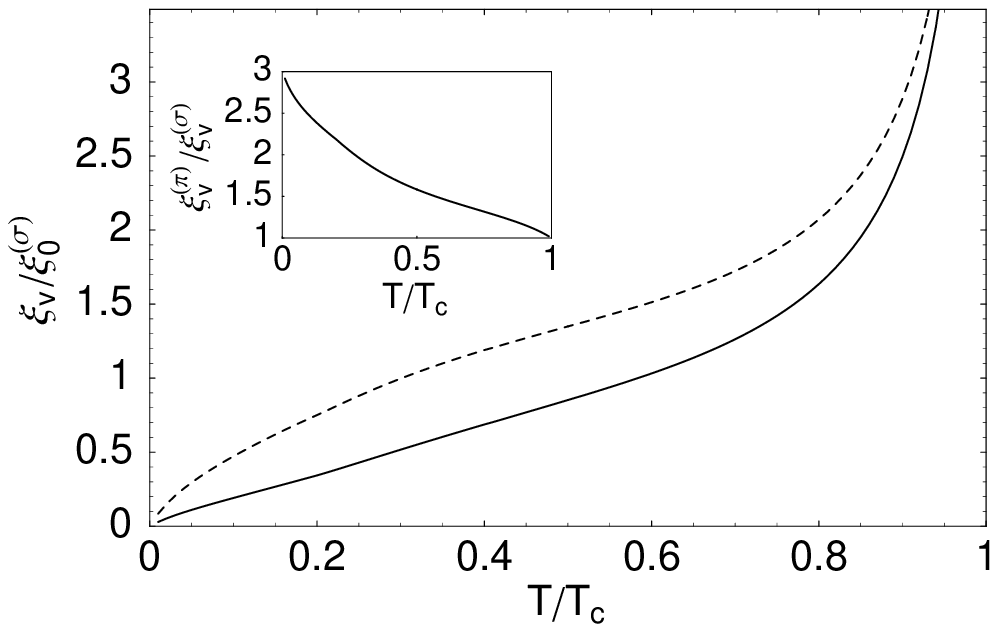}
\hspace*{0.1cm}
    \includegraphics[width=0.99\columnwidth,angle=0]{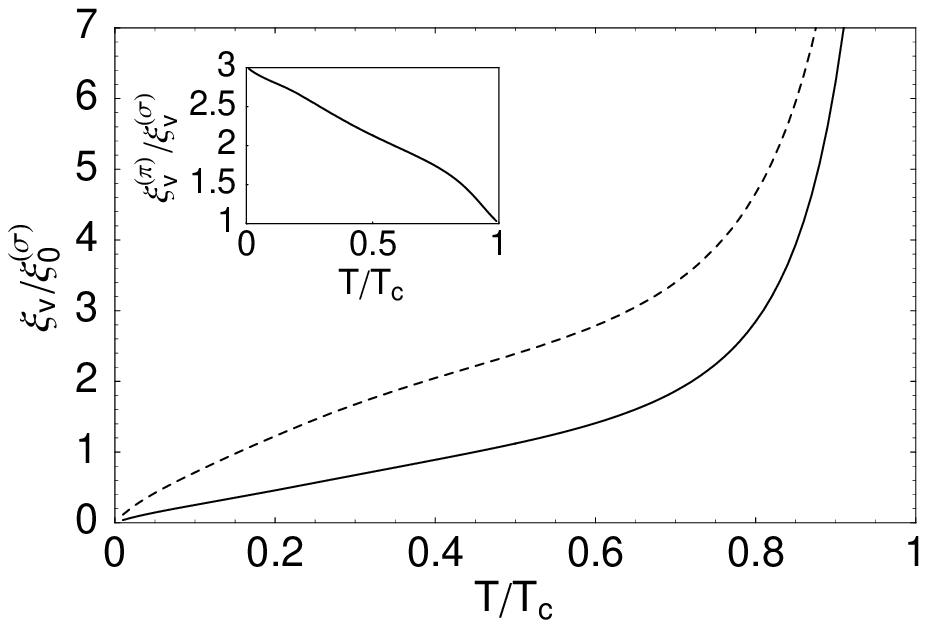}
    \caption{Temperature dependence of the vortex core radius 
	$\xi_V/\xi^{(\sigma)}$ in the $\sigma$ band (solid line) and
	in the $\pi$ band (dashed line) for different values of the
	coherence length ratio $\xi^{(\sigma)}/\xi^{(\pi)}$.
	(a) $\xi^{(\sigma)}/\xi^{(\pi)}=5$, (b) $\xi^{(\sigma)}/\xi^{(\pi)}=1$,
	and (c) $\xi^{(\sigma)}/\xi^{(\pi)}=0.2$. The insets show the
	temperature dependence of the ratio of the two vortex core radii
	$\xi^{(\pi)}_V/\xi^{(\sigma)}_V$.
    \label{fig1} }
  \end{center}
\end{figure} 

We conclude that the induced Kramer-Pesch effect reported earlier 
\cite{Gumann} exists independently of the diffusivity and coherence
length in the dirty band.

\end{document}